\documentstyle[twoside,fleqn,espcrc2]{article}

\newcommand{\gapproxeq}{\lower 
.7ex\hbox{$\;\stackrel{\textstyle >}{\sim}\;$}}
\newcommand{\lapproxeq}{\lower 
.7ex\hbox{$\;\stackrel{\textstyle <}{\sim}\;$}}

\newcommand{\AmS}{{\protect\the\textfont2
  A\kern-.1667em\lower.5ex\hbox{M}\kern-.125emS}}

\title{Issues for the Lattice in Hadron Spectroscopy}

\author{F E Close\address{Rutherford Appleton Laboratory, Chilton, Didcot, 
Oxon,
        OX11 OQX, England.}}

\begin{document}

\begin{abstract}
Lattice QCD predicts a rich spectroscopy of glueballs and $q\bar{q}-$glue hybrids.  
I compare these with data and assess the emerging empirical situation.  
Questions for the lattice are proposed.
\end{abstract}
\maketitle
\section{The Glue Landscape}

The collective behaviour of gluons in the strong interaction regime of QCD is 
one of the black holes of the standard model.  Lattice QCD and a range of QCD 
inspired models all suggest that the lightest glueballs occur in the 1-2 GeV region 
and that $q\bar{q}-$gluon hybrids occur around 2 GeV in mass, yet, a quarter of a 
century after QCD was developed,their existence 
is not yet definitely established.

The PDG\cite{pdg96} have extensive lists of meson properties.  These states fit 
well with a naive $q\bar{q}$ spectroscopy, at least up to around 2 GeV, without 
any clear evidence or need for extra glueballs.  While it may be the case that the 
emergence of a glueball spectrum above 2 GeV disturbs and complicates the 
$q\bar{q}$
 picture, the prediction that $J^{PC}=0^{++}$ occurs {\bf below} 2 GeV, and 
moreover around 1.6 GeV, ought to be testable.  For if we cannot establish the 
truth or otherwise of this, what hope do we have to establish the existence of 
strongly bound glue in general?

A reason why so much is known about $q\bar{q}$
 states is due to the emphasis through several decades on beams of $\pi, K,
p$  
made of quarks interacting with targets made of quarks which biased the 
production of $q\bar{q}$ states.  A further bias was the historical emphasis on 
detecting decays into charged particles such as $\pi^\pm, K^\pm$  and inability 
to detect easily $\eta$ or $\eta^\prime$ whose affinity for glue via
the $U_A(1)$ anomaly had been speculated.  Thus I 
suggested\cite{fec83,fec89} that 
glueball signals may be enhanced if one concentrated on production in a quark-
free environment and if one also studied decay channels involving $\eta\eta$ 
and $\eta\eta^\prime$ or even $\eta^\prime \eta^\prime$.

Three optimal production strategies were suggested.\\
i) \quad $\psi\rightarrow\gamma R$\cite{chan75} where, in pQCD, the 
$\psi(c\bar{c})\rightarrow \gamma gg \to \gamma R$. The intermediate $gg$ 
system favours states coupling to glue.\\
ii) \quad Central production, $pp\rightarrow p+ R+p$, where the resonance $R$ 
is produced displaced from the beam and target in rapidity as $s\rightarrow 
\infty$.  The idea is that diffractive processes are driven by gluons (the 
Pomeron)\cite{robson}.\\
iii) \quad $p\bar{p}$ annihilation at low energies, where 
$(qqq)+(\bar{q}\bar{q}\bar{q})\rightarrow \pi + gg$, and subsequent detection of 
etas (hence  $p\bar{p} \rightarrow \pi\eta\eta$ or  $\pi\eta\eta^\prime$) 
appeared to me to be a natural source of a $J^{PC}=0^{++}$ or $ 2^{++}$ in the 1-2 
GeV region.  The emergence of LEAR and especially the Crystal Barrel seemed 
particularly suitable\cite{fec83}.

Meanwhile, lattice QCD predictions for the mass of the lightest
scalar glueball were maturing. 
 To set the scene here is a summary of what lattice QCD predicts.

\subsection{Glueballs}

The SU(3) glueball spectrum for all $J^{PC}$ values at lattice spacings down to 
$\beta$ = 6.4 has been studied by ref.\cite{ukqcd}.
  The generic features include\\
i) \quad the lightest state has  $J^{PC} = 0^{++}$ with $m\sim 1.6 \pm$ 0.1 
GeV.\\
ii) \quad Below 3 GeV,  potentially accessible in $\psi\rightarrow\gamma G$, 
there are also $J^{PC} = 0^{-+}, 2^{++}$ and possibly also $J^{PC} = 2^{-+}$ and the 
spin exotic $J^{PC} = 1^{-+}$.\\
iii) There are signals for spin exotics, $J^{PC} = 0^{+-}, 1^{-+}$ and possibly $0^{--}$, 
below 4 GeV.

The predictions for the mass of the lightest scalar glueball have become
firmer at this conference\cite{morn}.

\subsection{Hybrids}

The spectrum of hybrid mesons produced by gluonic excitations in quenched 
QCD has been evaluated\cite{lacock,milc}.
  The most clear cut signal for a hybrid 
meson is to search for $J^{PC}$ quantum numbers not allowed in the quark 
model such as $J^{PC} = 1^{-+}, 0^{+-}$ and $2^{+-}$.  Light flavoured hybrids are 
predicted to contain $J^{PC} = 1^{-+}$ with mass 2.0 $\pm$ 0.2 
GeV\cite{lacock} (for the $s\bar{s}$ so the $n\bar{n}$ may be expected some
200-300 MeV lower)
 in agreement with $1.97 \pm 0.09 \pm 0.3$(syst) GeV\cite{milc}.
Heavy 
flavour hybrids include $c\bar{c} g$ at 4.19 $\pm$ 0.15 GeV\cite{lacock}.
Ref.\cite{milc} find $J^{PC} = 1^{-+}$ at $4.39 \pm 0.08 \pm 0.2$ GeV
with the $J^{PC} = 0^{+-}$ higher at $\sim 4.6$ GeV. The $b\bar{b}g$ hybrids are
predicted to be slightly above the $\Upsilon(4S)$ at $10.81 \pm 
0.25$GeV\cite{lacock}.

What evidence is there for any of these states being realised in Nature?  There 
have been some exciting developments in the last three years.  
These have been stimulated by the clear sighting of a scalar flavourless meson
$f_0(1500)$ that is, at least superficially, a glueball candidate and
of flavoured mesons $\pi(1800)$ and exotic $J^{PC}=1^{-+}$ with
properties consistent with being quark-gluon hybrids. The plan of the
talk will be to survey the primary glueball candidates $f_0(1500)$
and $f_{(J=0,2?)}(1710)$, then
address other possible states with the lattice and finally look at the hybrid 
meson scenario.

\section{The Scalar Glueball and the $f_0(1500)/f_J(1710)$}

A novel $J^{PC}=0^{++}, m \simeq$ 1.5 GeV, $\Gamma \sim$ 
100 MeV appears to be present in each of the three processes cited above as 
``glueball friendly". A state denoted $f_J(1710)$\cite{pdg96} (where
$J= 0$ or $2$) also is seen tantalisingly in $\psi \to \gamma f_J$,
in central production $pp \to pf_Jp$ and, recently, in
$p\bar{p}$ annihilation in flight.   In recent months two 
further processes have come into attention, each involving heavy flavours\\
a) \quad $D_s\rightarrow \pi\pi\pi$ where $D_s (\bar{s}c) \rightarrow W^+ 
gg \rightarrow \pi^+ (\pi\pi)$.  This can complement $D_s\rightarrow \pi 
(K\bar{K})$ and help establish the flavour content of $0^{++}$ mesons.\\
b) \quad $B\rightarrow K + (c\bar{c}) \rightarrow K gg \rightarrow K +$ light 
hadrons. This might provide an entree into the glueball sector and
possibly also $c\bar{c}g$ hybrid charmonium, up to $\sim 4$ GeV\cite{cdpv}.

A significant new result from the lattice\cite{weinprl} is that the two
body width of a scalar glueball is $\sim O(100)$ MeV and not $\sim
O(1000)$ MeV. In principle the glueball could have been extremely wide
and for practical purposes unobservable. The lattice shows that the
scalar glueball should be a reasonably sharp signal which is an
important guide in helping to eliminate candidates. The width for
decay of the  scalar glueball into pseudoscalar pairs was
predicted\cite{weinprl} to be $108 \pm 28$ MeV. The $f_0(1500)$
has $\Gamma_{tot} = 120 \pm 20$ MeV\cite{pdg96} with the decays into
pseudoscalar pairs comprising $\sim 60$ MeV of this. The $f_J(1710)$
has $\Gamma_{tot} = 140 \pm 12$ MeV, prominently in $0^-0^-$.
 The lattice prediction of the width
guides us towards these
states (if $f_J(1710)$ has $J=0$) but does not of itself discriminate between
them.

 I shall begin by reviewing the data on the $f_0$ (1500) and
$f_J(1710)$ in the three 
``classical" processes above.

\subsection{ $p\bar{p}$ annihilation}

A basic template in $pQCD$ is the $q\bar{q}g$ vertex as manifested in three-jet 
events, $e^+e^- \rightarrow q\bar{q}g$.  To form the simplest $(gg)_1$ gluonic 
system (where the subscript denotes the colour representation)
requires $(qq)_{\bar{3}} + (\bar{q}\bar{q})_3 \rightarrow (gg)_1$.  The 
only diquark ``beams" are contained within baryons and so we focus on 
$p\bar{p} \rightarrow \pi + (gg)_1 \rightarrow \pi (\eta\eta)$ and 
other relevant channels. Prior to LEAR there were few events on these
channels; post LEAR there are millions of events and it is this fact,
combined with the high resolution of the Crystal Barrel detector in particular,
that has revolutionised our picture of hadron spectroscopy in the critical
region of $\sim 1.5-2$ GeV.

If a $J^{PC}=0^{++}$ glueball really exists below 1.7 GeV, then the above suggests 
that it must be produced by $p\bar{p}$ at rest.  There seemed to be only two 
possible problems that might obscure it: 
(i) if the width is huge (recent lattice results 
suggest that this is unlikely\cite{weinprl,wein96}) (ii) if the $qqq$ and 
$\bar{q}\bar{q}\bar{q}$ rearrange themselves to form $3\pi$ as a background 
swamping any $\pi + G$ signal.  The vast data sample with high resolution from 
both Crystal Barrel and Obelix has overcome the latter problem.

There is one final problem.  
Established states such as $K_0$ (1430), $f_0$ (1370) 
(and subsequently $a_0$ (1450)\cite{pdg96}) and/or $q\bar{q}$ potential models 
imply that the $J^{PC}=0^{++}$ 
$^3P_0 q\bar{q}$  nonet is expected to occur in the 
vicinity of the primitive scalar glueball.  There will be mixing between the 
$n\bar{n}, G_0$ and 
$s\bar{s}$ states\cite{cafe95}.
  The resulting mixing pattern is rather robust when extended 
to a full 3x3 mixing\cite{wein96,cfl97,teper97} and has the following 
consequences.  Three physical states will occur that we refer to as $\phi_s, 
\phi_G, \phi_n$ which we identify as $f^n_0 \simeq 1370, f^{G/s}_0 \simeq 
1500, 
f^{s/G}_0 \simeq 1700$ 
(the latter state may be the $f_J$ (1710) or it may be shifted 
strongly by coupling to $K\bar{K}$ in $S-$wave and thereby be associated with 
$f_0$ (980). This is currently a hotly debated topic.

The empirical situation is as follows.

The  $f_0 (1500) $ is seen in  
$p\bar{p}\rightarrow\pi\eta\eta,\pi\eta\eta^\prime,\pi\pi\pi$ and $5\pi$.  
There are also emerging signs in $\pi K\bar{K}$.  The state clearly exists.  An 
extensive review of its properties in $\bar{p}p$ is given by Amsler\cite{am97}.  
See also ref.\cite{cafe95,wein96}. The suppression of $K\bar{K}$
and affinity for $\eta \eta$ and $\eta \eta'$ are consistent with the 
$G-q\bar{q}$ mixing\cite{cafe95}.  The mass and width are typically $m=1500\pm 
15, \Gamma = 120\pm 15$ MeV. Little is yet known about $f_J(1710)$ in 
$p\bar{p}$
as most data have been taken at rest where phase space kills $f_J(1710) + \pi$;
data in flight see a signal but a separation of $J=0,2$ remains to be made.

\subsection{ Production in $\psi\rightarrow\gamma R$}

Ref.\cite{cfl97} has used the measured radiative quarkonium rates and 
$\gamma\gamma$ decay widths to make quantitative estimates of the gluonic 
content of isosinglet mesons.  In essence, it describes the empirical $b.r.
 (\psi 
\rightarrow\gamma R)$ as a convolution between a piece calculable in pQCD 
and a residual non-perturbative unknown that is essentially the 
$b.r.(R\rightarrow gg$).  One may expect

$$
b(R[q\bar{q}] \rightarrow gg) = 0(\alpha^2_s) \simeq 0.1-0.2
$$
$$
b(R[G] \rightarrow gg) \simeq  0(1) 
$$
Thus knowledge of $b(R\rightarrow gg)$ would give quantitative information 
on the glueball content of a particular resonance.

Ultimately the proof of the pudding is in the eating.  A priori one could imagine 
that the extracted $b(gg)$ could be anything at all, greater than 100\% even, if the 
idea had no foundation.  However, when applied to known $q\bar{q}$ 
resonances one finds
$$
b.r.[f_2 (1270), f_2 (1525), f_1 (1280), f_1 (1420), \eta (1285)]
$$
$$
\simeq 0.2-0.3\simeq O(\alpha^2_s)
$$
This contrasts with
$br(\xi (2230))\simeq 1;  br(\eta (1410)) \simeq 0.9$ 
and $br(f_0 (1500)) \simeq 0.6$ 
(assuming that the signal in $\psi\rightarrow\gamma 4\pi$ is indeed 
$J^P=0^+$ as claimed by\cite{bugg96} and not $J^P=0^-$; this needs to be clarified 
at Beijing and at a Tau Charm Factory) and $br(f_J(1710)) \sim 0.8$ if $J=0$.
  
The $3 \times 3$ mixing schemes give a picture where
the branching ratios are expected to be (in units $10^{-3}$)\cite{cfl97}

$$
b(\psi\rightarrow \gamma  f_0 (1370) : \gamma  f_0 (1500) : \gamma  f_0 
(1700))
$$
\begin{equation}\label{psirad}
\simeq 0.2:0.5:1
\end{equation}
which appear to be consistent with an analysis of $\psi \to \gamma \pi \pi$
and $\gamma K\bar{K}$\cite{dun97}.
A general summary of glueball candidates, when confronted quantitatively
with $\psi \to \gamma R$ is as follows.

(i) The $f_0(1500)$\cite{cafe95,cfl97,bugg96}
 is probably produced at a rate too high to be a
$q\bar{q}$ state.  The average of world data suggests it is a
glueball-$q \bar{q}$ mixture.  

(ii) The $f_J(1710)$\cite{weing} where $J=0$ or 2\cite{pdg96}
 is produced at a rate which is consistent with
it being $q\bar{q}$, only if $J=2$.  If $J=0$, its production
rate is too high for it to be a pure $q\bar{q}$ state but is consistent
with it being a glueball or mixed $q \bar{q}$-glueball having a large
glueball component\cite{cfl97,dun97}. 

(iii) The $\xi(2230)$\cite{beijing}, whose width is $\sim 20$ MeV, is produced at 
a
rate too high to be a $q\bar{q}$ state for either $J=0$ or $2$.  If
$J=2$, it is consistent with being a glueball.  The assignment $J=0$
would require $Br(J/\psi \rightarrow \gamma \xi) \sim 3 ~10^{-4}$,
which already may be excluded. 

(iv) The enhancement once called $\eta(1440)$ has been resolved
into two states\cite{cfl97,suchung}.  The higher mass $\eta(1480)$ is dominantly
$s\bar{s}$ with some glue admixture, while the lower state
$\eta(1410)$ appears to have strong affinity for glue.

\subsection{ Suppression in $\gamma \gamma$}

By contrast, the $\gamma \gamma$ couplings are expected to be small
for a glueball and in the mixed states of refs.\cite{cafe95,wein96}
we expect the 
relative magnitudes of their production rates as follows\cite{cfl97}\\

\begin{equation}\label{gamma}
\Gamma (\gamma\gamma\rightarrow f^{nn}_0: f^G_0: f^s_0) \simeq 5: 0.1-
0.4:1
\end{equation}
If the empirical width $f_0 (1370)\rightarrow \gamma\gamma\simeq$ 2-5 keV 
then the $\Gamma (f_0 (1500)\rightarrow \gamma\gamma) \simeq$ 0.03-0.4
keV is the challenge for experiment. An optimal strategy for searching
for the $s\bar{s}$ member of the multiplet is to study
$\gamma \gamma \to f_0 \to K\bar{K}$ or $\eta\eta$ (also one should
study $D_s \to \pi K\bar{K}$, see later). Hints from LEP2 are that
$\gamma \gamma \to f_0(1500) \to \pi \pi$ implies that
$\Gamma(f_0(1500) \to \gamma \gamma) \leq 0.17$keV\cite{alison}

\subsection{ Scalar Mesons in $D_s$ decays}

$D_s$ decays can provide a direct window into the $0^{++}$ sector.  Qualitatively 
one expects a hierarchy
$$
\Gamma (D_s\rightarrow \pi (s\bar{s})) >
\Gamma (D_s\rightarrow \pi gg \to \pi G) > 
$$
$$
\Gamma (D_s\rightarrow \pi gg \to \pi (n\bar{n}))
$$
Empirically in $D_s\rightarrow \pi \pi n$ one sees $\pi f_0$ (980) as most 
prominent which is presumably due to the affinity of $f_0$ (980) for $s\bar{s}$, 
or $K\bar{K}$.  The empirical hierarchy from E687 is, roughly, for the 
$\pi\pi\pi$ channel
$$
b.r. (D_s\rightarrow \pi  f_0 (980): \pi f_0 (1500): \pi f_2 (1270) : \pi\rho) 
$$
$$
\simeq 0.65: 0.25: 0.08: 0.02
$$
The $\pi f_2 (1270)$ is probably driven by $\pi gg \rightarrow \pi (n\bar{n})$ 
whereas $\pi\rho$ comes from annihilation $D_s\rightarrow W\rightarrow 
\pi_g$\cite{lipkin96}. The observation of $f_0$ (1500) (the fit requires $M$= 1475, 
$\Gamma$ = 100 which is well consistent with the $f_0$ (1500)) is interesting as 
is its production rate.  An analysis of $D_s\rightarrow \pi (gg)\rightarrow \pi 
R$ along the lines of ref.\cite{cfl97} could be informative. 
It is now important also to 
study $D_s\rightarrow \pi K\bar{K}$ and $\pi\eta\eta$ in order to access the 
$s\bar{s}$ number of the $J^{PC} = 0^{++}$ nonet and help to elucidate the nature 
of $f_0$ (1500) and $f_J$ (1720). In particular it will be central to
determining if $f_0(980)$ is the only state below $\sim 1800$ MeV that
couples strongly to $s\bar{s}$: this could have major implications
for establishing the $s\bar{s}$ content of the $^3P_0$ nonet.

\subsection{ Central Production}

In $pp\rightarrow p(4\pi)p$ and $p(2\pi)p$, WA91 and WA102\cite{wa102}
 see a scalar signal in the 1500 
MeV region.  However this cannot be immediately claimed as support for a 
glueball since, in the $4\pi$ spectrum at least, the most prominent structure is 
the $f_1$ (1285), a well established $q\bar{q}$ state.  Thus we infer that both 
$q\bar{q}$ and $G$ can be produced and that there may be interference between 
the $^3P_0  q\bar{q} f_0$ (1370) and a glueball candidate $f_0$ (1500).  Indeed 
there does appear to be non-trivial interference since the mass and/or width are 
not simply identified with those of the $f_0$ (1500) $(m\sim 1500, \Gamma\sim 
100 MeV)$. WA91 found
$$
i)\mbox{ in} 2\pi : m = 1500 \pm 30 MeV~ \Gamma = 200\pm 30 MeV
$$
$$
ii)\mbox{ in} 4\pi : m = 1445\pm 5 MeV~ \Gamma = 65\pm 10 MeV
$$

Clearly central production is not simply a glueball factory and we need to look 
into it more critically.

\section{How to Make Glueballs when Protons Collide}

Feynman imagined two protons colliding in their c.o.m.  Each proton consists of 
partons carrying longitudinal momentum functions $0\leq x \leq 1$.  
Those with $x\rightarrow 1$ are clearly right or left movers in the c.o.m. and 
unambiguously belong to the right or left moving proton respectively.  However 
those with $x\rightarrow 0$ in either left or right mover are (near) at rest in the  
c.o.m. and can transfer from left to right moving proton without disturbing the 
protons' wavefunctions.  In Feynman's picture it is the exchange of these 
$x\rightarrow 0$ partons that is responsible for elastic scattering.

Measurement of the parton distribution functions show that as  $x\rightarrow 0$ 
the partons are most likely to be gluons.  Thus we may envision elastic scattering 
at high energies as mediated by the exchange of gluons in an overall colour 
singlet configuration.

Now consider the exchange of partons (gluons) with $x_{1,2} \neq 0$.
The two gluon beams may 
fuse to produce a meson whose overall momenta will be
$$
p_L \equiv (x_1-x_2)P ; p_T \equiv q_{1T} + q_{2T}
$$
and for $x_1 \simeq x_2$ the mass will be 
$$
M^2_R\simeq 4x_1x_2 P^2 - P^2_T \simeq  4x_1x_2 P^2
$$
(where the experimental conditions are typically $4P^2\equiv s\simeq$ 900 
GeV$^2$ and the 
$M^2_R$ range of interest is 1-4 GeV$^2$).  To form an exclusive meson 
the relative momentum of the incident gluon ``beams", namely $(x_1+x_2) 
P\simeq 2x_1 P \simeq M_R$, must be redistributed so that the constituents of 
the produced meson have a {\bf small} relative momentum $\lapproxeq 
0(\Lambda_{QCD})$.  Thus considerable rescattering will be required, especially 
at large $M_R$, and so one expects that the gluons may fuse directly into a 
glueball or may undergo $gg\rightarrow q\bar{q}$ and form a quarkonium.

The question that Kirk and I addressed is whether one may alter the kinematic 
conditions and enhance the glueball production relative to $q\bar{q}$.  We 
suggested that one study the spectrum as a function of\cite{kirk97}
$$
dp_T=|\vec{q}_{T1} - \vec{q}_{T2}|
$$
since, for a given $M_R$, when $|dp_T|$ is large, more rescattering is required 
(in the transverse direction) than when $|dp_T|$ is small.  Thus for $|dp_T|$ 
small one may hope that glueballs may be more favoured; for $|dp_T|$ large, by 
contrast, the $q\bar{q}$ may be relatively enhanced.
Refs.\cite{kirk97,wa10297} and found that all undisputed $q\bar{q}$
mesons are suppressed at small $dp_T$ whereas glueball candidates are 
enhanced.
Specifically, 

(i)when $dp_T >$ 0.5 GeV/c the $q\bar{q}$ states $f_1$ (1285) and $f_1$ (1420) are 
clearly seen in the $K\bar{K}\pi$ channel; when 0.2 $< 
dp_T <$ 0.5 GeV the $q\bar{q}$ are still visible, though rather less prominent, 
whereas for $dp_T <$ 0.2 GeV they have all but disappeared into the background.

(ii)The $f_2(1270)$ and $f_2(1525)$ show similar behaviour in the $\pi \pi$
and $K\bar{K}$ channels respectively: they
 only become apparent as $dp_T$ increases.  

(iii) In the $K\bar{K}$ spectrum it is also noticeable that the $q\bar{q}$ 
$f_2$ (1525) is produced dominantly at high $dp_T$ whereas the 
enigmatic $f_J$ (1710) is 
produced dominantly at low $dp_T$.

(iv) The $4\pi$ channel is particularly rich. At large $dp_T$ the $q\bar{q}$
$f_1(1285)$, $\eta_2(1700)$ and possibly $f_4(2040)$ are seen with the $f_1(1285)$
particularly sharp. However, when $dp_T <$ 0.2 GeV the 
$f_1$ (1285), a $q\bar{q}$ state, has essentially disappeared 
as do the $\eta_2$ and $f_4$ while the  $f_0$ (1500) 
and an enigmatic $f_2$ (1900) structure have become 
more clear.  These surviving structures 
have been identified as glueball candidates: the $f_0$ (1500)  is motivated by 
lattice QCD 
while the  $f_2$ (1900) is noted to have the right mass to lie on the 
Pomeron trajectory\cite{prl}. 

Thus we have a tantalising situation in central production of mesons.  We have 
stumbled upon a remarkable empirical feature that does not appear to have been 
noticed previously.  Although its extraction via the $dp_T$ cut was inspired by 
intuitive arguments we have no simple dynamical explanation.  
An  interesting question is whether similar phenomena occur in $ep\rightarrow 
eRp$ or $e^+e^-\rightarrow e^+ Re^-$. These can be investigated at HERA 
or in $e^+e^-$ colliders if the outgoing beams are tagged\cite{fec97}.

The $f_0(1500)$ shares
features expected for a glueball that is mixed with the nearby
isoscalar members of the $^3P_0$ $q\bar{q}$ nonet. In particular 
ref\cite{cafe95} noted
that this gives a destructive interference between $s\bar{s}$ and $n\bar{n}$
mixing whereby the $K\bar{K}$ decays are suppressed.
The properties of the $f_J(1710)$ become central to completing the glueball
picture.  If the $f_J(1710)$
proves to have $J=2$, then it is not a candidate for the ground state
glueball and the $f_0(1500)$ will be essentially unchallenged.
On the other hand, if the $f_J(1710)$ has $J=0$ it becomes a potentially
interesting glueball candidate.  Indeed, Sexton, Vaccarino and
Weingarten\cite{weinprl} argue that $f_{J=0}(1710)$ should be
identified with the ground state glueball, based on its similarity in
mass and decay properties to the state seen in their lattice
simulation. 
The prominent scalar
$f_0(1500)$ was originally interpreted by them\cite{weinprl} 
as the $s\bar{s}$ member of the scalar nonet, however  this identification
does not fit easily with the small $K\bar{K}$ branching ratio and 
the dominant decays to pions.

Whereas the spin of the $f_J(1710)$ remains undetermined, 
it is now clearly established that there are scalar mesons $f_0(1370)$
and $f_0(1500)$ \cite{pdg96} which couple to $\pi \pi$ and $K\bar{K}$
and so must be allowed for in any analysis of this mass region.

The presence of $f_0(1370), a_0(1450), K_0(1430)$ reinforce the expectation
that a $q\bar{q}$ $^3P_0$ nonet is in the $O(1.3 - 1.7)$GeV mass region.
It is therefore extremely likely that an `ideal' glueball at $\sim 1.6$GeV
\cite{teper97}
will be degenerate with one or other of the $^3P_0$ states given that the widths
of the latter are $O($hundreds MeV). This has
 not been allowed for in any lattice simulation so far.

The emerging concensus is that the scalar glueball is not a singleton
and that there are significant gluonic 
components in the nearby $n\bar{n}$ and $s\bar{s}$ states. Ref.\cite{cafe95}
proposed that ``if the $f_J(1710)$ is confirmed to have a $J=0$ component in
$K\bar{K}$ but not in $\pi \pi$, this could be a viable candidate for a
$G_0-s\bar{s}$ mixture, completing the scalar meson system built on the
glueball and the quarkonium nonet".

Recently Weingarten\cite{wein96} has proposed
what at first sight appears to be a different mixing scheme 
based on estimates for the mass of the $s\bar{s}$
scalar state in the quenched approximation. Whereas ref\cite{cafe95} supposed 
that the ideal glueball lies within the nonet, ref\cite{wein96} supposed
it to lie above the nonet. I shall now start with the general expressions of
ref\cite{cafe95} and compare the two schemes. This will reveal some rather
general common features.

\section{Three-State Mixings}
\label{3mixing}
\hspace*{2em}
An interesting possibility is that three $f_0$'s in the $1.4-1.7$ GeV 
region are admixtures of the three isosinglet states $gg$, $s\bar s$, and
$n\bar n$\cite{cafe95}.  
At leading order in the glueball-$q \bar{q}$ mixing, ref\cite{cafe95}
obtained 
\begin{eqnarray}
\label{mixing}
N_G|G\rangle = |G_0\rangle + \xi ( \sqrt{2} |n\bar{n}\rangle + \omega 
|s\bar{s}
\rangle) \nonumber \\
N_s|\Psi_s\rangle = |s\bar{s}\rangle - \xi \omega |G_0\rangle \nonumber 
\\
N_n|\Psi_n\rangle = |n\bar{n}\rangle - \xi  \sqrt{2} |G_0\rangle
\end{eqnarray}
where the $N_i$ are appropriate normalisation factors, $\omega \equiv
\frac{E(G_0) - E(d\bar{d})}{E(G_0) - E(s\bar{s})}$ and the mixing parameter
$\xi \equiv \frac{\langle d\bar{d}|V|G_0\rangle}{E(G_0) - E(d\bar{d})}$. The
analysis of ref\cite{cfl97}
suggests that the $gg \to q\bar{q}$ mixing amplitude manifested
in $\psi \to \gamma R(q\bar{q})$ is $O(\alpha_s)$, so that qualitatively
$\xi \sim O(\alpha_s) \sim 0.5$. Such a magnitude implies significant mixing
in eq.(\ref{mixing}) and is better generalised to a $3 \times 3$ mixing matrix.
Ref.\cite{wein96} defines this to be
$$
\begin{array}{c c c c}
m_G^0 & z & \sqrt{2} z \\
z & m_s^0 & 0  \\
\sqrt{2} z & 0 & m_n^0 \\
\end{array}
$$ 
where
$z \equiv \xi \times (E(G_0) - E(d\bar{d}))$ in the notation of 
ref.\cite{cafe95}.

Mixing based on lattice glueball masses lead to two classes of solution
of immediate interest:  

\noindent (i)$\omega \leq 0$, corresponding to $G_0$ in the midst
of the nonet\cite{cafe95} 

\noindent (ii)$\omega > 1$, corresponding to $G_0$ above the
$q\bar{q}$ members of the nonet\cite{wein96}. 

We shall denote the three mass eigenstates by $R_i$ with $R_1=f_0(1370)$,
$R_2=f_0(1500)$ and $R_3=f_0(1710)$, and the three isosinglet states 
$\phi_i$ with $\phi_1=n\bar n$, $\phi_2=s\bar s$ and $\phi_3=gg$ so
that $R_i=f_{ij}\phi_i$.  

There are indications from lattice QCD that the scalar $s\bar{s}$ state,
in the quenched approximation, may lie lower than the scalar glueball
\cite{lacock96,wein96}. Weingarten\cite{wein96} has constructed a
mixing model based on this scenario. The input ``bare" masses are
$m_n^0 = 1450; m_s^0 = 1516; m_G^0 = 1642$ and the mixing strength
$z \equiv \xi \times (E(G_0) - E(d\bar{d})) = 72$ MeV. The
resulting mixtures are
$$
\begin{array}{c c c c}
&f_{i1}^{(n)} & f_{i2}^{(s)} & f_{i3}^{(G)}\\
f_0(1370) & 0.87 &  0.25 & -0.43\\
f_0(1500) & -0.36 &  0.91 & -0.22\\
f_0(1710) & 0.34 & 0.33 & 0.88\\
\end{array}
$$ 

It is suggested, but not demonstrated, that the decays of the
$f_0(1500)$ involve significant destructive interference between its
gluonic and $s\bar{s}$ components whereby the $K\bar{K}$ suppression and
$2\pi$, $4\pi$ enhancements are explained.

Recent data on the decay $f_0(1500) \to
K\bar{K}$\cite{landua} may be interpreted within the scheme of
ref\cite{cafe95} as being consistent with the $G_0$ lying between
$n\bar{n}$ and $s\bar{s}$ such that the parameter $\omega \sim
-2$. (In this case the $\eta \eta$ 
production is driven by the gluonic component of
the wavefunction almost entirely,see ref\cite{cafe95}).
If for illustration we adopt $\xi
=0.5 \sim \alpha_s$, the resulting mixing amplitudes are 
$$
\begin{array}{c c c c}
&f_{i1}^{(n)} & f_{i2}^{(s)} & f_{i3}^{(G)} \\
f_0(1370) & 0.86 &  0.13 & -0.50\\
f_0(1500) & 0.43 & - 0.61 & 0.61\\
f_0(1710) & 0.22 & 0.76 & 0.60\\
\end{array}
$$

The solutions for the lowest mass state in the two schemes
are similar, as are the relative
phases and qualitative importance of the $G$ component in the high
mass state.  Both solutions exhibit destructive interference between
the $n\bar{n}$ and  $s\bar{s}$ flavours for the middle state. 

This
parallelism is not a coincidence.
A general feature of this three way mixing is that in the limit of
strong mixing the central state tends towards flavour octet with the
outer (heaviest and lightest) states being orthogonal mixtures of
glueball and flavour singlet, namely 

$$
\begin{array}{c c}
f_0(1370) & |q\bar{q}(\bf{1})\rangle - |G\rangle \\
f_0(1500) & |q\bar{q}(\bf{8})\rangle + \epsilon |G\rangle \\
f_0(1710) & |q\bar{q}(\bf{1})\rangle + |G\rangle \\
\end{array}
$$
where $\epsilon \sim \xi^{-1} \to 0$.

In short, the glueball has leaked away maximally
to the outer states even in the case (ref\cite{cafe95}) where the bare glueball
(zero mixing) was in the middle of the nonet to start with. The leakage into
the outer states becomes significant once the mixing strength (off diagonal
 term in the mass matrix) becomes comparable to the mass gap between glueball
and $q\bar{q}$ states (i.e. either $\xi \geq 1$ or $\xi \omega \geq 1$).
 Even in the zero width approximation of ref\cite{cafe95}
this tends to be the case and when one allows for the widths being of 
$O(100)$MeV while the nonet masses and glueball mass are spread over only
a few hundred MeV, it is apparent that there will be considerable leakage
from the glueball into the $q\bar{q}$ nonet. It is for this reason,
 {\it inter alia}, that the output of refs\cite{cafe95} and \cite{wein96}
are rather similar. While this similarity
may make it hard to distinguish between
them, it does enable data to eliminate the general idea should their
common implications fail empirically.

If we make the simplifying assumption that the photons couple to the
$n\bar{n}$ and  $s\bar{s}$ in direct proportion to the respective
$e_i^2$ (i.e. we ignore mass effects and any differences between the
$n\bar{n}$ and $s\bar{s}$ wavefunctions), then the corresponding two
photon widths can be written in terms of these mixing coefficients:
\begin{equation}\label{mixings}
\Gamma(R_i)=|f_{i1}\frac {5}{9\sqrt{2}}+f_{i2}\frac {1}{9}|^2 \Gamma,
\end{equation}
where $\Gamma$ is the $\gamma\gamma$ width for a $q\bar q$ system
with $e_q=1$.  One can use eq. (\ref{mixings}) to evaluate the
relative strength of the two photon widths for the three $f_0$ 
states with the input of the mixing coefficients\cite{cfl97}.
  If we ignore
mass dependent effects, these lead to the results in eqn.\ref{gamma}.
We anticipate $f_0(1500) \to \gamma
\gamma \sim 0.3 \pm 0.2$ keV \cite{cafe95} or $\sim 0.1$ keV \cite{wein96}.
Both schemes imply $\Gamma(f_0(1710) \to \gamma \gamma) =
O(1)$ keV. 

This relative ordering of $\gamma \gamma$ widths is a common feature
of mixings for all initial configurations for which the bare glueball
does not lie nearly degenerate to the $n\bar{n}$ state.  As such, it
is a robust test of the general idea of $n\bar{n}$ and $s\bar{s}$
mixing with a lattice motivated glueball.  If, say, the $\gamma
\gamma$ width of the $f_0(1710)$ were to be smaller than the
$f_0(1500)$, or comparable to or greater than the $f_0(1370)$, then
the general hypothesis of significant three state mixing with a
lattice glueball would be disproven. The corollary is that qualitative
agreement may be used to begin isolating in detail the mixing pattern. 

The production of these states in $\psi \to \gamma f_0$ also shares some 
common
features in that $f_0(1710)$ production is predicted to dominate. The
analysis of ref.\cite{cfl97} predicts that
\begin{equation}
br(J/\psi \to \gamma \Sigma f_0) \geq (1.5 \pm 0.6) \times 10^{-3}.
\end{equation}
In \cite{cafe95} the $q\bar{q}$ admixture in the $f_0(1500)$ is nearly
pure flavour octet and hence decouples from $gg$. This leaves the
strength of  $br(J/\psi \to \gamma f_0(1500))$ driven entirely by its
$gg$ component at about $40\%$ of the
pure glueball strength. This leads to eqn.\ref{psirad} which
appears to be consistent with the mean of the 
world data (\cite{bugg96,dun97,cfl97}).

Thus, in conclusion, both these mixing schemes imply a similar hierachy
of strengths in $\gamma \gamma$ production which may be used as a 
test of the general idea of three state mixing between glueball and
a nearby nonet. Prominent production of $J/\psi \to \gamma f_0(1710)$
is also a common feature.  When the experimental situation clarifies
on the $J/\psi \to \gamma f_0$ branching fractions, we may be able 
to distinguish between the case where the glueball
lies within a nonet, ref\cite{cafe95}, or above the $s\bar{s}$ member,
ref\cite{wein96}. 

In the former case this $G_0 - q\bar{q}$ mixing
gives a destructive interference between $s\bar{s}$ and $n\bar{n}$
whereby decays into $K\bar{K}$ are suppressed.
However, even in the case where the $s\bar{s}$ lies below the $G_0$
we expect that there will be
$K\bar{K}$ destructive effects due to mixing not only
with the $s\bar{s}$ that lies
below $G_0$ (as in ref.\cite{wein96}) but also 
with a radially excited $n\bar{n}$
lying above it (not considered in ref.\cite{wein96}). Unless $G_0$ mixing with
the radial state is much suppressed, this will give a similar pattern to that
of ref.\cite{cafe95} though with more model dependence due to the differing
spatial wavefunctions for the two nonets.

\section{The Hybrid Candidates}

When the gluon degrees of freedom are excited in the presence of  $q\bar{q}$ one 
has so called ``hybrid" states.  In lattice QCD and/or models one expects these 
states (denoted $\pi_g, D_g, \psi_g$ to mean gluonic excitation with overall 
flavour quantum numbers of a $\pi, D$ or $c\bar{c}$ etc) to occur with masses 
$\pi_g \sim$ 1.8 ~GeV, $D_g \sim$ 3 GeV, $\psi_g \sim$ 4 GeV.  
There are
tantalising sightings of an emerging spectroscopy as I shall now review.

It is well known that hybrid mesons can have $J^{PC}$ quantum numbers 
 in combinations such as $0^{--},0^{+-},
1^{-+}, 2^{+-}$ etc. which are unavailable to conventional mesons and as
such provide a potentially sharp signature. 

It was noted in ref.\cite{kokoski85} and confirmed in ref.\cite{cp95}
that the best opportunity for isolating exotic hybrids appears
to be in the $1^{-+}$ wave where, for the I=1 state with mass around 2 GeV,
partial widths are typically

\begin{equation}
\label{bnlwidth}
 \pi b_1 : \pi f_1 : \pi \rho \;
= \; 170 \; MeV : 60 \; MeV : 10 \; MeV
\end{equation}
The narrow $f_1(1285)$ provides a useful tag for the 
$1^{-+} \rightarrow \pi f_1$ and ref.\cite{lee94} has recently reported a signal
in $\pi^- p \rightarrow (\pi f_1) p$ at around 2.0 GeV
that appears to have a resonant phase.

Note
the prediction  that the $\pi \rho$ channel is 
not negligible relative to the signal channel
$\pi f_1$   
thereby resolving the puzzle of the production
mechanism that was commented upon in ref. \cite{lee94}.
This state may also have been sighted in photoproduction \cite{utk}
with $M=1750$ and may be the $X(1775)$ of the Data Tables, ref.\cite{pdg96}.
There has also been recent claim for a possible exotic $J^{PC}=1^{-+}$
around 1.4GeV decaying into $\pi \eta$\cite{e852} which is also
reported from LEAR\cite{learexotic}. There is also a signal
around 1.6GeV in $\pi \eta'$\cite{vesexotic}. The experimental situation
here needs to be settled before the lattice predictions are confronted
directly but there does seem a likelihood that we can anticipate the
emergence of a hybrid spectroscopy to be compared with the lattice and
QCD inspired models.

A recent development
is the realisation that even when hybrid and conventional mesons
have the {\bf same} $J^{PC}$ quantum numbers, they may still be distinguished
\cite{cp95} due to their different internal structures which give rise to
characteristic selection rules. When conventional quantum numbers such as
$0^{-+}$ are analysed on the lattice, it is found that the conventional 
mesons, such as the $\pi$, have considerable signal\cite{mich}. In order to 
separate 
the genuine $\pi$ from the hybrid signal on the lattice
it would be interesting to exploit the 
different spin content of the hybrid and ground state configurations.

Turning to the $0^{-+}$ wave, 
the VES Collaboration at Protvino and BNL E852 both see
a clear $0^{-+}$ signal in diffractive  $\pi N \to \pi \pi \pi N$.
\cite{had95}. Its mass and decays typify those
expected for a hybrid: $M \approx 1790$ MeV, $\Gamma \approx 200$ MeV
in the $(L=0)$ + $(L=1)$ $\bar{q}q$
 channels $\pi^- + f_0; \; K^- + K^*_0, \; K {( K \pi )}_S $ with no
corresponding strong signal in the kinematically allowed $L=0$ two body 
channels $\pi + \rho; \; K + K^*$. This confirms the earlier sighting
by Bellini et al\cite{bellini}, listed in the Particle Data group\cite{pdg96}
as $\pi(1770)$.

The resonance also appears to couple as strongly to
the enigmatic $f_0(980)$ as it does to $f_0(1300)$,
which was commented upon with some surprise in ref. \cite{had95}.
This may be natural for a hybrid at this mass due to the
predicted dominant $KK_0^*$ channel which will feed
the $(KK\pi)_S$ (as observed \cite{had95}) and hence the channel
$\pi f_0(980)$ through the strong affinity of $K\bar{K} \rightarrow f_0(980)$.
Thus the overall expectations for hybrid $0^{-+}$ are in line with
the data of ref.\cite{had95}. 

  This $\pi_g$
%\cite{cp95,isgur} 
is, accidentally, 
degenerate with the $D$ and so may affect the Cabibbo suppressed decays of the 
latter\cite{lipkin96}.  A comparison between decays of $\pi_g$\cite{ves} and the 
Cabibbo suppressed decays of $D$ as measured by E687 collaboration at Fermilab, 
show some parallels.  Is is possible that this accidental degeneracy could give a 
non-perturbative enhancement of CP violation in the $D$ 
system\cite{lipkin96,buccella}.  To test further the idea that the $\pi_g$ affects 
$D$ decays, one should search in $D$ decays for channels that have shown up in 
$\pi_g$ decay.  For example: if $\pi_g$ is a guide, then $D\rightarrow 
K\bar{K}\pi$ in $S-$wave will be significant and  $D\rightarrow  \eta\eta\pi^-$ 
should occur at about 50\% intensity of $\pi^+\pi^-\pi^-$.  Finally, the glueball 
candidate $f_0$ (1500) should occur in  $D\rightarrow \pi f_0$ (1500).

This is an instructive example of where light flavour spectroscopy can
affect the dynamics of heavy flavour decays. In the final section I shall turn 
to heavy flavours. There have been interessting developments both in the
lattice, for hybrid charnmonium, and in phenomenology related to $B$ decays
as a possible source of the hybrid charmonia and of heavy glueballs.

\subsection{ Hybrid Charmonium and missing charm}

The decay $B\rightarrow K+ (c\bar{c})$ produces the $(c\bar{c})$ dominantly in 
a colour 8.  It may thus be an entree into the hybrid charmonium sector.  
Furthermore the lightest such states are predicted by lattice QCD\cite{hyb}
 to occur around 
4.1 $\pm$ 0.1 GeV whereby $B\rightarrow K+ \psi_g$ may be favoured.  
Finally, in models, such states decay strongly into $DD^{**}$, for which the 
threshold is 4.3 GeV, and not into $D\bar{D}, DD^*, D^*\bar{D}^*$.  
Consequently their preferred decays could be\\
a)  cascades to $\psi,\chi,\eta_c,h_c$ + light hadrons\\
b) decays to light hadrons via resonant glueballs, namely $(c\bar{c}_8 
g\rightarrow g^* g\rightarrow$ light hadrons.  As such this could be an entree 
into the spectrum of glueballs, predicted by lattice QCD to lie between 2-4 GeV 
and including exotic states $J^{PC}= 0^{+-}, 1^{-+}, 2^{+-}$.  This has been 
suggested by ref.\cite{cdpv}

In summary: the lattice predictions that the lightest glueball is
a scalar merge tantalisinngly with the discovery of an enigmatic
$f_0(1500)$ and with the possibility that the $f_J(1710)$ contains a
significant $J=0$ component. 
They are produced in the right mass region, according
to the lattice, and in the right processes, according to intuition
developed from knowledge of hadron dynamics. Important now is to 
understand the dynamics behind the central production kinematic filter
that appears to distinguish the glueball candidates from established
$q\bar{q}$ mesons. Finally, to complete the scalar nonet, we need to establish 
the $s\bar{s}$ member. The $f_0(980)$ may be the remnant of this state, shifted 
to $K\bar{K}$ threshold by its $S$-wave coupling to mesons, or there may be
a state to be established around $1700-1800$ MeV: the spin of the
$f_J(1710)$ and its relation to the $f_0(1500)$ is critical in this
respect. The decays $D_s \to
\pi K \bar{K}$ and the production via $\gamma \gamma \to f_0 \to K 
\bar{K}$
in contrast to $\gamma \gamma \to f_0 \to \pi \pi$ promise the most
direct resolution of this question. Finally, the decays of $B \to K +$ light
hadrons may reveal intersting new dynamics up to $\sim 4$ GeV in mass.

\end{document}